\documentclass[9pt]{extarticle}
\usepackage{spconf,amsmath,graphicx}
\usepackage{color,booktabs}
\usepackage{subfig}
\usepackage[ruled, linesnumbered]{algorithm2e}
\usepackage{amsfonts}
\usepackage{hyperref}

\title{DiffRoll: Diffusion-based Generative Music Transcription\\with Unsupervised Pretraining Capability}
\name{
\begin{tabular}{@{}c@{}}
Kin Wai Cheuk$^{1,2,*}$\thanks{$^*$Work done during internship at Sony} \qquad 
Ryosuke Sawata$^{3}$ \qquad 
Toshimitsu Uesaka$^{3}$  \qquad
Naoki Murata$^{3}$ \\
Naoya Takahashi$^{3}$ \qquad 
Shusuke Takahashi$^{3}$ \qquad
Dorien Herremans$^{1}$ \qquad
Yuki Mitsufuji$^{3}$
\end{tabular}}
\address{$^1$Singapore University of Technology and Design, Singapore\\
$^2$Agency for Science, Technology and Research, Singapore\\
$^3$Sony Group Corporation, Tokyo, Japan}
\begin{document}

%
\maketitle
\begin{abstract}
In this paper we propose a novel generative approach, DiffRoll, to tackle automatic music transcription (AMT).
Instead of treating AMT as a discriminative task in which the model is trained to convert spectrograms into piano rolls, we think of it as a conditional generative task where we train our model to generate realistic looking piano rolls from pure Gaussian noise conditioned on spectrograms.
This new AMT formulation enables DiffRoll to transcribe, generate and even inpaint music. Due to the classifier-free nature, DiffRoll is also able to be trained on unpaired datasets where only piano rolls are available. 
Our experiments show that DiffRoll 
outperforms its discriminative counterpart by 19 percentage points (ppt.) and our ablation studies also indicate that it outperforms similar existing methods by 4.8 ppt.
Source code and demonstration are available at \url{https://sony.github.io/DiffRoll/}.
\end{abstract}
\begin{keywords}
Automatic Music Transcription, Music Information Retrieval, Signal Processing, Diffusion, Generative Model, Unsupervised Pretraining
\end{keywords}
\section{Introduction}
\label{sec:intro}

Automatic Music Transcription (AMT) has typically been treated as a discriminative task~\cite{nam2011classification} in which the frequency bins of spectrograms are projected onto 88 notes (from A0 to C8) in posteriorgrams. There have been attempts at modeling AMT using Bayesian inference~\cite{cemgil2004bayesian} and probabilistic models~\cite{berg2014unsupervised}, however, these models lack the generative power found in variational autoencoders (VAEs)~\cite{roberts2018musicvae} and generative adversarial networks (GANs)~\cite{engel2018gansynth,dong2018musegan}.

In addition, the majority of existing state-of-the-art (SOTA) AMT models are fully supervised~\cite{Hawthorne2017OnsetsAF,gardner2021mt3,10.1109/TASLP.2021.3121991,cheuk2022jointist,wei2022hppnet}.
While there have been attempts to develop weakly-supervised~\cite{pmlr-v162-maman22a} and semi-supervised AMT models~\cite{wu2022semi, cheuk2021reconvat}, unsupervised deep learning AMT is still an under-explored direction and thus currently limited to drum transcription~\cite{DBLP:conf/ismir/ChoiC19} and non-deep learning based approaches~\cite{berg2014unsupervised, smaragdis2003non}. In this paper, we have devised a novel formulation for AMT called ``DiffRoll'', in which the power of diffusion is leveraged, making it a generative model capable of generating new piano rolls with the potential of unsupervised pretraining. 

\section{Related Work}
\label{sec:related_work}

Diffusion model in deep learning inspired by the concept of non-equilibrium thermodynamics~\cite{diffusion-sohl-dickstein15}, has shown promising results in generating highly-fidelity audio samples~\cite{kong2020diffwave,chen2020wavegrad}. Subsequent improvements have led to better diffusion methods, for example, better forward process~\cite{ho2020denoising}, new regularization~\cite{lai2022_fokker_planck}, faster sampling speed~\cite{song2020denoising}, and discrete diffusion for binary data~\cite{chen2022analog}. In the field of vision, conditional diffusion such as DALL-E~\cite{ramesh2021zero} enables generated images to be controlled by an input text, resulting in SOTA text-to-image performance.

Diffusion is now also being used for music-related tasks such as music generation~\cite{mittal2021_musicgen_diffusion} and music synthesis~\cite{hawthorne2022multi}. These tasks are intuitively considered as generative and hence it is reasonable to model them using diffusion. AMT, on the other hand, has typically been considered a discriminative task in which an AMT model classifies the notes present in the spectrograms frame-by-frame. Even though there have been attempts~\cite{gardner2021mt3,yan2021skipping} to model AMT as a token-based task inspired by natural language processing, these approaches are still discriminative. AMT, however, can be modeled as a generative task. The spectrograms can be considered as the conditions, and the piano rolls can be considered as the image to be generated. In other words, we can think of AMT as a piano roll generation task, which is the reverse process of music synthesis~\cite{hawthorne2022multi}. In this paper, we take a novel perspective by modeling AMT as a generative task, with the source code and demonstration available\footnote{\url{https://sony.github.io/DiffRoll/}}.

\section{Proposed Method}
\begin{figure}[htb]
  \centering
  \includegraphics[width=8.5cm]{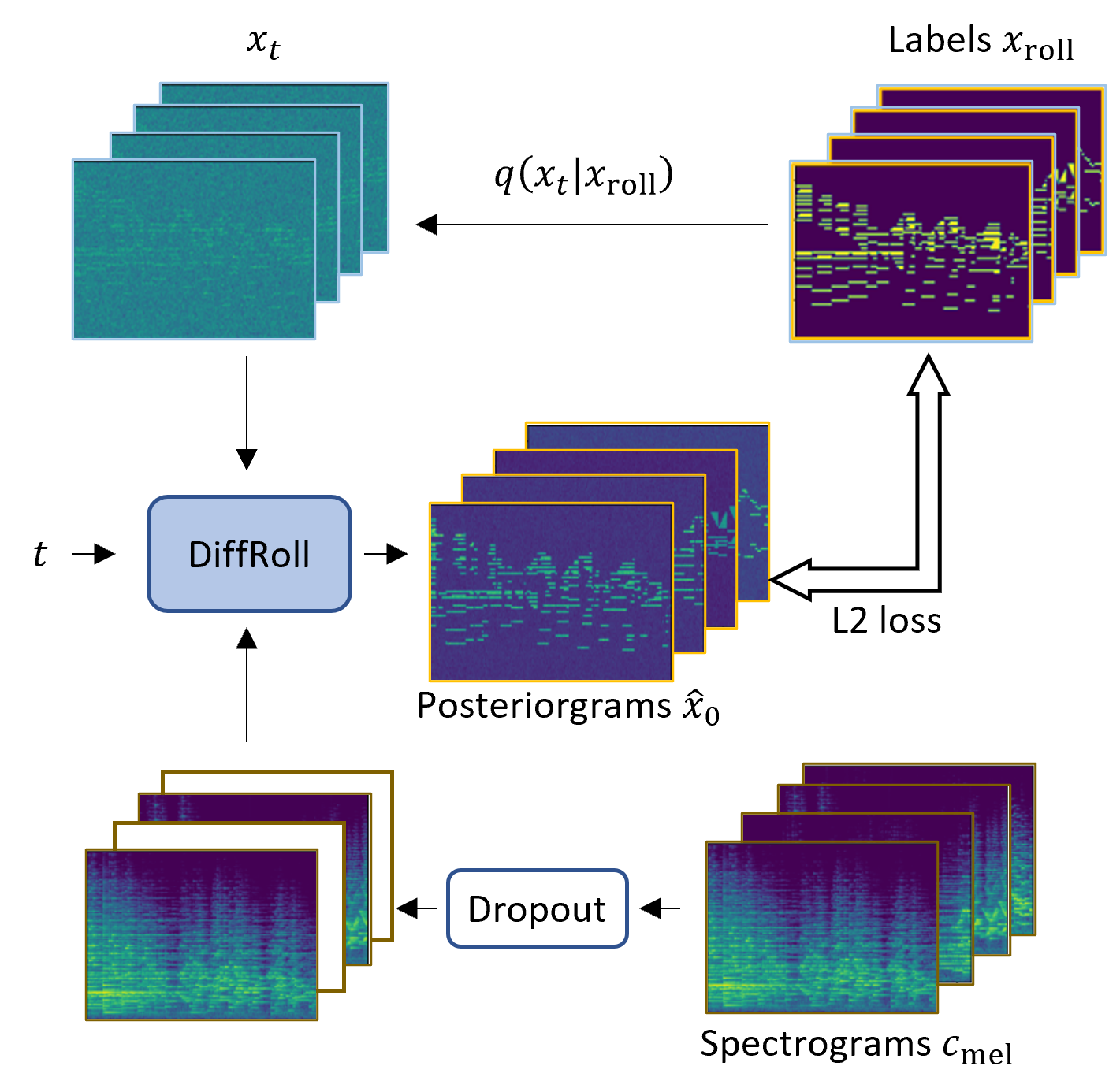}
\caption{Training phase of DiffRoll. Within each batch, several spectrograms are randomly converted to -1 via the dropout layer.}
\vspace{-5mm}
\label{fig:model}
\end{figure}

DiffRoll, as shown in Fig.~\ref{fig:model}, is designed to convert Gaussian noise $x_t$ into a posteriorgram $\hat{x}_0$ conditioned on spectrogram $c_\text{mel}$. To handle binary piano rolls $x_\text{roll}\in\{0,1\}$ during training, we cast them into $[0,1]$, similar to Analog Bits~\cite{chen2022analog}. During sampling, the posteriorgram is binarized back into piano roll with a threshold of $0.5$ and then exported as a MIDI file.

The DiffRoll model architecture is inspired by DiffWave~\cite{kong2020diffwave} which is also a 1D convolutional model. In DiffWave, $x_t$ is a 1-channel tensor, while in DiffRoll we consider $x_t$ as an 88-channel tensor with a dimension of $(B,88,\tau)$, where $B$ is the batch size and $\tau$ is the number of frames in the piano roll. As in DiffWave, $x_t$ is projected into a tensor with shape $(B,512,\tau)$ via a 1D convolutional layer with a kernel size of $1$. A total of 15 residual layers with the same design as in DiffWave are used, where the output of the previous residual layer is added to the input of the next layer. A diffusion time $t$, i.e. a batch of integers with shape $(B,1)$, is projected into $(B,512)$ and then broadcasted to $(B,512,\tau)$ such that it can be added to the input of each residual layer. Note that to not confuse diffusion time $t$ with the time dimension of the spectrograms or piano rolls $\tau$, $t$ and $\tau$ are completely different from each other. To condition the posteriorgram generation $\hat{x}_0$, there are 1D convolutional layers projecting the mel spectrograms from $(B,229,\tau)$ to $(B,512,\tau)$ such that they can be added to the output of the residual layers. Finally, a tensor with shape $(B,512,\tau)$ is projected back to a tensor with shape $(B,88,\tau)$ as the generated posteriorgram $\hat{x}_0$.

During \textbf{forward process}, denoted as $q(x_t|x_0)$ in Fig.~\ref{fig:model}, noisy piano roll $x_t$ is directly sampled from clean piano roll $x_\text{roll}$ using Eq.~\eqref{eq:forward}. 
\begin{equation}
\label{eq:forward}
    x_t = \sqrt{\bar{\alpha}_t}x_\text{roll}+\sqrt{1-\bar{\alpha}_t}\epsilon,
\end{equation}
where $\epsilon\sim\mathcal{N}\left(\textbf{0},\textbf{I}\right)$ is sampled from a standard Gaussian distribution, $\bar{\alpha}_t=\prod_{s=1}^t\alpha_s$, and $\alpha_0=1$. We used linear noise scheduling in which $\alpha_t\in[0.9999,0.98]$.

During \textbf{training}, model $f_\theta(x_t, t, c_\text{mel})$ with weight $\theta$ directly outputs $\hat{x}_0$ given noisy piano roll $x_t$, diffusion step $t$, and mel spectrogram $c_\text{mel}$, where $t$ is uniformly sampled from $[1,2,\dots,T]$. We set $T=200$ as preliminary experiments demonstrated that it yields the best result. To enable DiffRoll to transcribe and generate music, we need both a \textbf{conditional} as well as an \textbf{unconditional} model as defined in Eq.~\eqref{eq:cfg}.
\begin{equation}
\label{eq:cfg}
\hat{x}_0(\cdot) = 
    \begin{cases} f_\theta(x_t, t, c_\text{mel}), & \mbox{conditional} 
    \\ f_\theta(x_t, t, -{1}), &\mbox{unconditional} 
    \end{cases},
\end{equation}
Although both models should produce the same output $\hat{x}_0(\cdot)$, we use $\hat{x}_0(c_\text{mel})$ and $\hat{x}_0(-1)$ to differentiate their outputs. To train both models at the same time without introducing extra parameters, we use the concept of Classifier-Free Guidance (CFG)~\cite{ho2021classifierfree}. More specifically a dropout layer as shown in Fig.~\ref{fig:model}, is used to randomly mask $c_\text{mel}$ in each batch by $-1$ with probability $p$. We chose $-1$ as the dropout value due to the fact that $0$ corresponds to silence in $c_\text{mel}\in[0,1]$ and we want to avoid confusing the model during training. The model is trained to minimize the L2 loss between model output $\hat{x}_0(\cdot)$ and ground truth label $x_\text{roll}$, as shown in Eq.~\eqref{eq:loss}, similar to previous approaches~\cite{kawar2022denoising,salimans2022progressive}. A full explanation of this loss is provided in the supplementary material\footnotemark[1].


\begin{equation}
\label{eq:loss}
L = \|x_\text{roll}-\hat{x}_0(\cdot)\|^2
\end{equation}

Although model output $\hat{x}_0$ can be easily converted into $x_{t-1}$ via $q(x_{t-1}|\hat{x}_0,x_t)$ (line 4-6 of Algorithm~\ref{alg:sampling}), $x_{t-1}$ is not used during training. The usage of $x_{t-1}$ is discussed further in the next paragraph.

\begin{algorithm}
    \DontPrintSemicolon
    $x_T\sim\mathcal{N}\left(\textbf{0},\textbf{I}\right)$\;    
    \For{$t=T,\dots,1$}{
        $\epsilon\sim\mathcal{N}\left(\textbf{0},\textbf{I}\right)$ if $t>1$, else $\epsilon=0$\;    
        $\hat{x}_0 = (1+w)\hat{x}_0(c_\text{mel}) - w\hat{x}_0(-1)$\; \label{eq:x_0}
        $\hat{\epsilon}_\theta^t=(x_t-\sqrt{\bar{\alpha}_t}\hat{x}_0)(\sqrt{1-\bar{\alpha}_t})^{-1}$\;\label{eq:epsilon}
        $x_{t-1}=\sqrt{\bar{\alpha}_{t-1}}\hat{x}_0+\sqrt{1-\bar{\alpha}_{t-1}-\sigma_t^2}\hat{\epsilon}_\theta^t+\sigma_t\epsilon$\; \label{eq:sampling}
        }
    \textbf{return} $x_0$
    \caption{Sampling}
    \label{alg:sampling}    
\end{algorithm}

\begin{figure}[t!]
  \centerline{\includegraphics[width=\columnwidth]{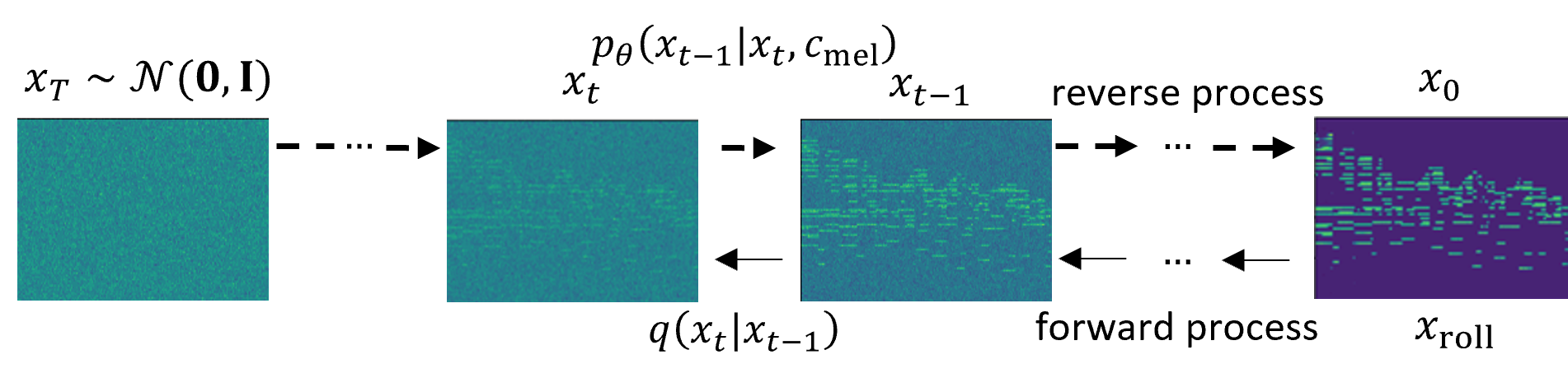}}
\caption{Diffusion process of DiffRoll. Forward process converts $x_\text{roll}$ into $x_t$ for $1\le t\le T$ via a Markov process. Reverse process $p_\theta(x_{t-1}|x_t,c_\text{mel})$ is performed iteratively until $x_0$ is reached. A threshold of 0.5 is applied to $x_0$ to obtain the predicted piano roll.}
\label{fig:trajectory}
\end{figure}

We designed Algorithm~\ref{alg:sampling} for \textbf{sampling} as shown in Fig.~\ref{fig:trajectory}.
First, $x_T$ ($T=200$ in our experiments) is sampled from a standard Gaussian distribution $\mathcal{N}(\textbf{0},\textbf{I})$. Then Eq.~\eqref{eq:cfg} is used to obtain conditional output $\hat{x}_0(c_\text{mel})$ and unconditional output $\hat{x}_0(-1)$. Next, a weighting $w$ is applied to both $\hat{x}_0(c_\text{mel})$ and $\hat{x}_0(-1)$ on line~\ref{eq:x_0} of Algorithm~\ref{alg:sampling} to obtain $\hat{x}_0$. Finally, $\hat{x}_0$ is used on line~\ref{eq:sampling} to obtain $x_{t-1}$, which is denoted as $q(x_{t-1}|\hat{x}_0,x_t)$. Note that line~\ref{eq:sampling} requires predicted noise $\hat{\epsilon}_\theta^t$ at diffusion time step $t$, whereas our model directly predicts $\hat{x}_0$. Therefore, we need line~\ref{eq:epsilon} of Algorithm~\ref{alg:sampling} to calculate $\hat{\epsilon}_\theta^t$ from $\hat{x}_0$. We denote this reverse process step as $p_\theta(x_{t-1}|x_t,c_\text{mel})$ in Fig.~\ref{fig:trajectory}.
The reverse process is performed iteratively until $x_0$ is reached. We experimented with both $\sigma_t=\sqrt{(1-\bar{\alpha}_{t-1})/(1-\bar{\alpha}_t)}\sqrt{1-\alpha_t}$ (DDPM~\cite{ho2020denoising}) and $\sigma_t=0$ (DDIM~\cite{song2020denoising}) and found that DDIM yields slightly worse transcription accuracy. Therefore, we used $\sigma_t$ for DDPM in our experiments. 




\section{Experiment}
\label{sec:exp}

The training split and test split of the MAESTRO dataset~\cite{hawthorne2018enabling} were used to train and test our models. During training, we used the validation split to monitor the loss. For simplicity, we ignored the velocity information of the piano roll. Following previous work~\cite{hawthorne2018enabling,cheuk2020impact}, we downsampled all of the audio clips to $16$kHz. The nnAudio~\cite{cheuk2019nnaudio} library was used to extract mel spectrograms with $229$ mel bins ranging from $0$ to $8,000$Hz, with a window size of $2,048$ samples from the audio segments ($327,680$ samples). The magnitude of the mel spectrograms was compressed by taking the natural logarithm and then normalizing it so that $c_\text{mel}\in[0,1]$. Although the original diffusion model was designed to work with $x_t\in[-1,1]$ $\forall t\in[0,T]$, our experimental results show that there are no extra benefits in normalizing the piano rolls into $x_\text{roll}\in\{-1,1\}$. DiffRoll is able to learn to output $\hat{x}_0\in[0,1]$ matching the original piano roll values $\{0,1\}$. Therefore, we kept $x_\text{roll}\in\{0,1\}$ throughout our experiments. We trained our models for 2,500 epochs on this dataset.

During the evaluation, we reported only the note-wise transcription metric (F1 score) since previous work has shown that the frame-wise transcription metric does not reflect perceptual accuracy well~\cite{cheuk2021IJCNN_revisit}. We binarized the transcription $x_0$ and extracted note objects from it. Each note object $\{\rho,\tau_\text{on},\tau_\text{off}\}$ contained three attributes: pitch $\rho$, onset time $\tau_\text{on}$, and offset time $\tau_\text{off}$. Following previous work~\cite{hawthorne2018enabling,cheuk2021IJCNN_revisit}, we used a music information retrieval library, \texttt{mir\_eval}\footnote{\url{https://github.com/craffel/mir\_eval}}, to calculate the note-wise transcription F1 score between the predicted note objects and the ground truth note objects. The predicted pitch $\hat{\rho}$ needs to match the ground pitch $\rho$ and the difference between the predicted and ground truth onset $\lvert\hat{\tau}_\text{on}-\tau_\text{on} \lvert$ needs to be within 50ms for the prediction to be considered correct.

\begin{table}[htb]
\centering
\caption{DiffRoll F1 scores with different $w$ values and dilation settings when $k=9$.}\label{tab:dilation}
\begin{tabular}{c|c|c|c|c}
 & \multicolumn{4}{c}{$w$} \\\toprule
Dilation      & \textbf{$0$} & \textbf{$0.1$} & \textbf{$0.5$} & \textbf{$1$}  \\\toprule
$[1,1,1,1,\dots,]$     &\textbf{ 74.9}         & \textbf{75.9}           & \textbf{78.1}          &       \textbf{77.7}                    \\
$[1,2,1,2,\dots,]$     & 74.0         & 75.3           & 77.1           &       76.9                      \\
$[1,2,4,1,\dots,]$   & 73.3         & 74.0           & 76.2           & 75.8                  \\
$[1,2,4,8,\dots,]$ & 72.2         & 74.0           & 75.9           & 76.1 \\\bottomrule 

\end{tabular}
\end{table}

\subsection{DiffRoll versus Discriminative AMT}
\label{sec:diffroll}
\begin{figure}[tb]
  \centering
  \centerline{\includegraphics[width=8cm]{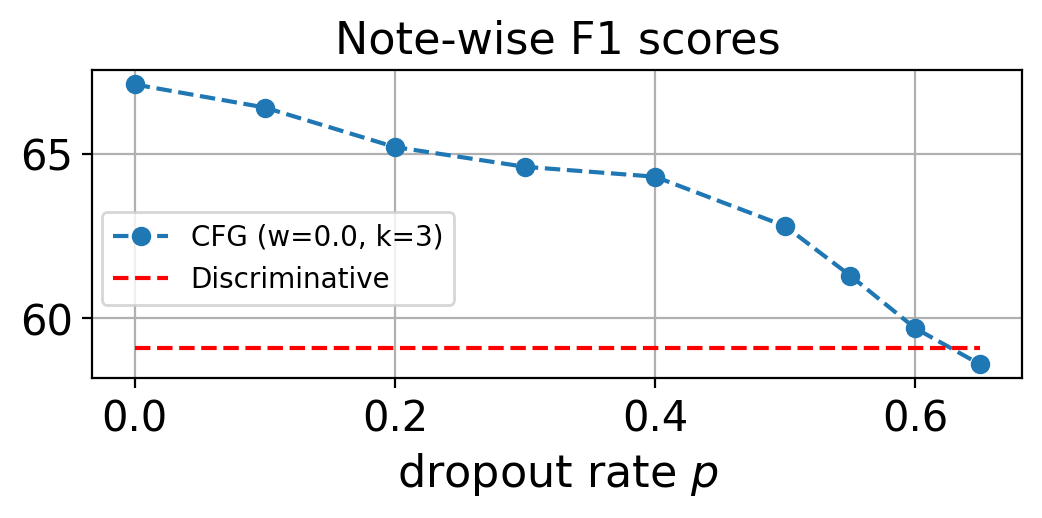}}
\caption{Result of DiffRoll ablation study with different values of dropout rate $p$. Baseline was discriminative version of DiffRoll, which directly predicts $x_0$.}
\vspace{-3mm}
\label{fig:F1_p}
\end{figure}

\begin{figure}[tb]
  \centering
  \centerline{\includegraphics[width=8.5cm]{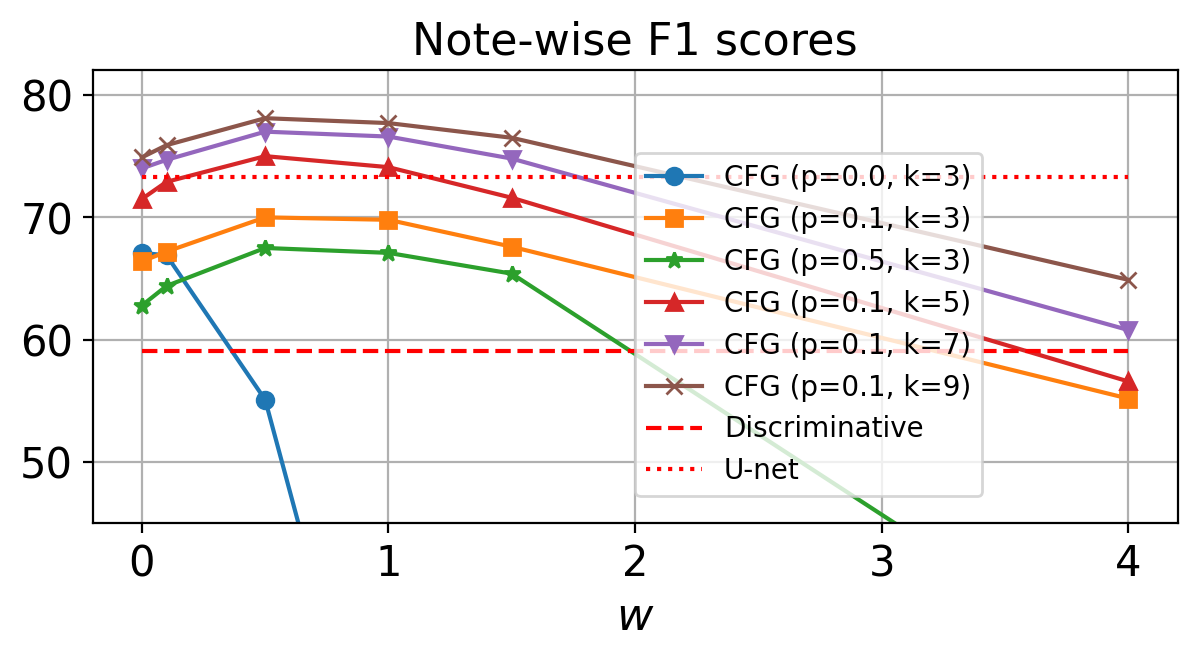}}
\caption{Results of CFG ablation study with different values of sampling weight~$w$. Discriminative AMT model and U-net model~\cite{cheuk2021icpr} were the baseline models.}
\label{fig:F1_w}
\end{figure}

In this experiment, we trained a traditional discriminative AMT model as the baseline. To ensure a fair comparison, we kept the model architecture exactly the same as that of the DiffRoll model, but used a zero tensor as the $x_t$, and a constant $t=1$ to directly predict the posteriorgram $\hat{x}_0$. After training, the discriminative model learned to ignore the $x_t$ and $t$ as they contained useless information for the task. This discriminative baseline (convolutional kernel size $k=3$) had an F1 score of $59.1$ as shown in both Figs.~\ref{fig:F1_p} and~\ref{fig:F1_w}. Fig.~\ref{fig:F1_p} also shows DiffRoll's performance for different values of dropout rate $p$. When $p=0$, the model was equivalent to standard diffusion without CFG. The transcription F1 score decreased as $p$ increased. This was expected since DiffRoll has a lower probability of obtaining paired data when $p$ is large. Although the best F1 score ($67.1$) was achieved with $p=0$, the F1 score rapidly decreased well below $20$ once a generative component was introduced ($w>0$ on line~\ref{eq:x_0} of Algorithm~\ref{alg:sampling}) as shown in Fig.~\ref{fig:F1_w}. The results of the ablation study in Fig.~\ref{fig:F1_w} also reveal that when $p>0$ and $w=0.5$, DiffRoll reached its peak performance of $70.0$. This experiment was done using kernel size $k=3$. 

Since piano rolls have long term structures, we also investigated different receptive field sizes by changing the kernel size ($k$) for DiffRoll. As we can see from Fig.~\ref{fig:F1_w}, the performance increased (from $70$ to $78.1$) as we increased $k$ from $3$ to $9$, which shows that a larger receptive field is beneficial for improving transcription accuracy. The degree of improvement, however, decreased as $k$ increased, so we stopped at $k=9$ and used this for the experiments described below.

In DiffWave~\cite{kong2020diffwave}, dilation is used to create a suitable receptive field to capture the long-term information present in the waveform. We experimented with different dilation settings (Table~\ref{tab:dilation}) for DiffRoll. Dilation factor $[1,2,1,2,\dots]$ indicates that the 1st and 2nd layers have dilation factors of $1$ and $2$, respectively; the same pattern holds for the 3rd and 4th layers. However, we observed no improvement in the F1 score when changing the dilation. On the contrary, the larger the dilation factor, the worse the F1 score. Therefore, we believe that a kernel size of $9$ without dilation has a large enough receptive field for the AMT task.

This experiment shows that modeling AMT as a conditioned generative task is possible and beneficial. A full animation for the generative trajectory of the transcription process is available in the supplementary material\footnotemark[1]. A DiffRoll model with kernel size $k=9$, dropout rate $p=0.1$, sampling weight $w=0.5$, and without any dilation had the best performing model and outperformed the discriminative counterpart ($59.1$ F1 scores) by 19 percentage points (ppt.) and even the U-net model~\cite{cheuk2021icpr} ($73.3$ F1 scores) by 4.8 ppt.. The model could not outperform current SOTA AMT models~\cite{hawthorne2018enabling,10.1109/TASLP.2021.3121991}, likely because we have not yet incorporated onset and offset information during training. We will enhance DiffRoll performance by incorporating onset and offset information as our future work.

\subsection{Unsupervised Pretraining}
\label{sec:unsupervised}

DiffRoll also supports unsupervised pretraining. Hence, it is possible to pretrain DiffRoll with piano rolls only (without paired audio). When audio data is not available, we can simply set $c_\text{mel}=-1$ to train only the unconditional model. To demonstrate this ability, we use MAESTRO~\cite{hawthorne2018enabling} as the unpaired piano roll dataset (967 unpaired piano rolls), and MAPS~\cite{emiya2010maps} as the paired dataset (139 audio clips paired with piano rolls). We first pretrained DiffRoll for 2,500 epochs using only the piano rolls in the training split of the MAESTRO dataset. We then continued training DiffRoll with the training split of the MAPS dataset for 10,000 epochs using paired audio clips and piano rolls. We removed overlapping samples between the training split and the test split from the MAPS dataset following previous work~\cite{Sigtia2015AnEN}. There are two ways to train on the MAPS dataset. One is to set dropout $p=0.1$ such that the 139 audio and piano roll data paired in the MAPS training split are used to train both the conditional and unconditional parts of DiffRoll. The other is to load both the MAPS and MAESTRO datasets, and set $p=0$ for the former and set $p=1$ for the latter. We denote this dropout scheme as $p=0+1$ in Table~\ref{tab:unsupervised}.

\begin{table}[h]
\centering
\caption{The performance of DiffRoll when pretrained (denoted as `pre-DiffRoll') using unpaired piano rolls (extracted from MAESTRO). `Discriminative' is the discriminative counterpart of DiffRoll, and `DiffRoll' was trained without unpaired pretraining. `ReconVAT' is an existing semi-supervised AMT model.}\label{tab:unsupervised}
\begin{tabular}{lcc}
\toprule
& \multicolumn{2}{c}{Note-wise F1 score}\\
Model      & $w=0$  & $w=0.5$\\\midrule
Discriminative & \multicolumn{2}{c}{50.6}\\
DiffRoll ($p=0.1$)  & 55.9 &     57.7 \\
pre-DiffRoll ($p=0.1$)& 59.0 & 62.6\\
pre-DiffRoll ($p=0+1$) & 60.5 &  63.7\\
ReconVAT~\cite{cheuk2021reconvat} & \multicolumn{2}{c}{64.0}\\\bottomrule
\end{tabular}
\end{table}

As shown in Table~\ref{tab:unsupervised}, the pretrained DiffRoll (pre-DiffRoll) achieved an F1 score $4.9$ points higher than the DiffRoll trained on the MAPS dataset from scratch when $p=0.1$ and $w=0.5$. When training scheme $p=0+1$ was used, the F1 score further improved by 1.1 to 63.7. Although DiffRoll was unable to outperform the SOTA semi-supervised AMT model ReconVAT~\cite{cheuk2021reconvat}, its score was competitive (63.7 v.s. 64.0). We believe that the model architecture of DiffRoll can be further improved and leave such improvement for future work. The same as in Section~\ref{sec:diffroll}, performance with $w=0.5$ was always better than with $w=0$ during sampling.

\subsection{Generation}
Unlike traditional discriminative AMT models, the proposed DiffRoll model is generative. In other words, it is capable of generating new piano rolls which can be considered as a form of music generation. To generate new music during sampling, we can reuse the same model we trained as described above, and set $w=-1$ during sampling such that the conditional term on line~\ref{eq:x_0} of Algorithm~\ref{alg:sampling} is cancelled out, leaving only the unconditional term. A full animation of the generation trajectory of $x_t$ via line~\ref{eq:sampling} of Algorithm~\ref{alg:sampling} during generation is available in the supplementary material\footnotemark[1]. In each iteration of line~\ref{eq:sampling} of Algorithm~\ref{alg:sampling}, DiffRoll makes $x_t$ less and less noisy. This can be seen from the animation: as $t$ approaches 0, the value range of $x_t$ narrows to around $[0,1]$ even though the output of DiffRoll is unbounded. Although the generated piano roll is far from perfect, we can still observe patterns of chords. This shows that our generative model successfully captures the data distribution for piano rolls $p(x_\text{roll})$.

\subsection{Inpainting}

\begin{figure}[th]
  \centerline{\includegraphics[width=\linewidth]{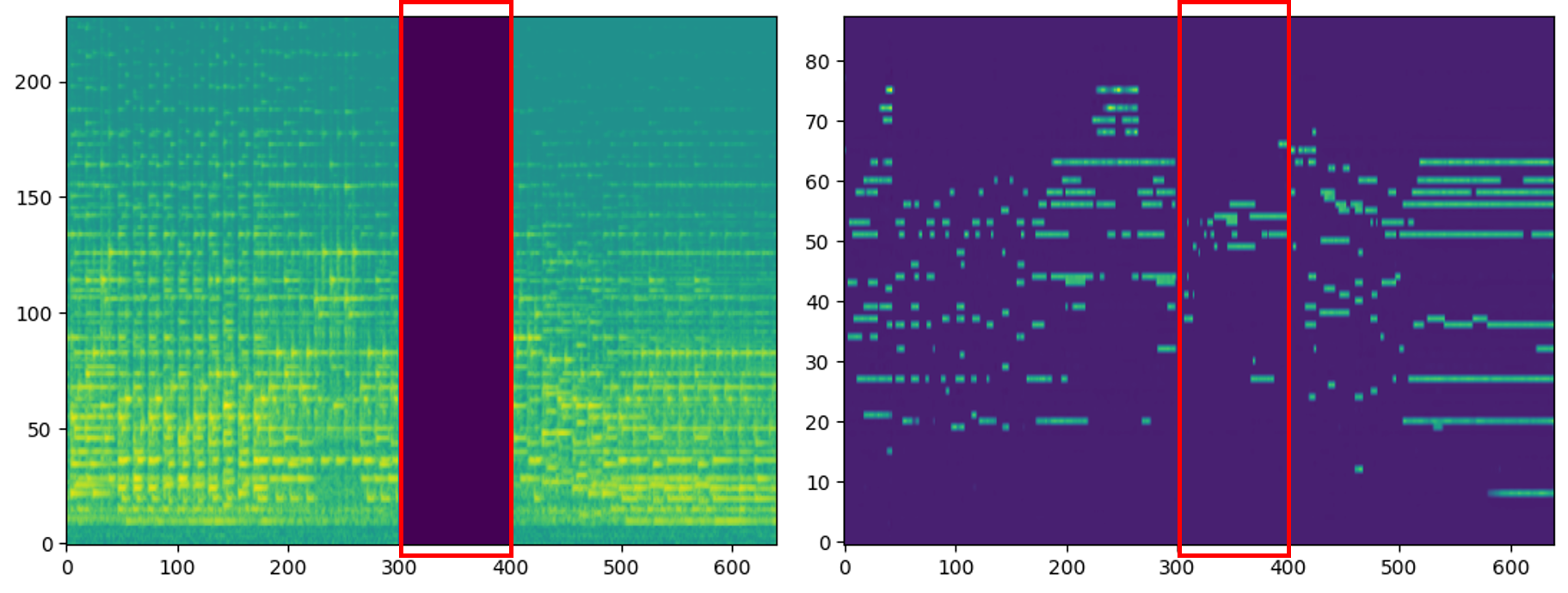}}
\caption{Piano roll inpainting when part of $c_\text{mel}$ is masked by -1 (indicated by red rectangles). DiffRoll generates notes in masked regions and transcribes notes in unmasked regions. More examples are available in the demo.}
\label{fig:inpainting}
\end{figure}

In the previous subsections, we described using DiffRoll for either transcription or generation. In this subsection, we briefly explore the inpainting capability of DiffRoll in which the input spectrograms $c_\text{mel}$ are partially masked with $-1$ during sampling. We use the same sampling scheme as on line~\ref{eq:sampling} of Algorithm~\ref{alg:sampling}. As shown in Fig.~\ref{fig:inpainting}, DiffRoll performs transcription when $c_\text{mel}>0$, and generation when $c_\text{mel}=-1$. More inpainting examples are available in the supplementary material\footnotemark[1]. Since inpainting is beyond the scope of this paper, in-depth exploration is left for future research.

\section{Conclusion}\label{sec:conclusion}

We proposed a novel generative automatic music transcription (AMT) model, DiffRoll. This model is capable of transcribing, generating, and even inpainting music. The generative property of DiffRoll enables it to be pretrained on unpaired piano rolls. Experiments showed that this unsupervised pretraining improves model performance when the amount of labelled data is limited. DiffRoll outperformed its discriminative counterpart by $7.1$ in terms of the F1 score on the MAPS dataset. When unsupervised pretraining was used (the MAESTRO dataset), the F1 score was further improved by $6$. Although DiffRoll did not outperform state-of-the-art AMT models, the results presented provide a successful proof of concept for how diffusion can be used for AMT. There are number of ways in which DiffRoll can be improved. Future work will include exploring the possibility of predicting onset and offset locations using diffusion, which has been shown to improve AMT accuracy. We will also investigate replacing Gaussian diffusion with discrete diffusion for binary piano rolls. This novel formulation of AMT should attract more research into the generative approach to AMT. The source code and generation examples are available in the supplementary material\footnotemark[1].





\vfill\pagebreak

\bibliographystyle{IEEE}
{\footnotesize
\bibliography{string}}

\end{document}